\documentstyle[preprint,eqsecnum,aps,axodraw,rotate]{revtex}

\begin{document}

\preprint{$
\begin{array}{l}
\mbox{KEK--TH--353}\\[-3mm]
\mbox{KEK preprint 92--202} \\[-3mm]
\mbox{February 1993} \\[2cm]
\end{array}$}

\title{Consequences of a Possible Di-Gamma Resonace at TRISTAN}

\author{\large K.\,Hagiwara, S.\,Matsumoto and M.\,Tanaka}
\address{Theory Group, KEK, Tsukuba, Ibaraki, 305, Japan}
\date{\today}
\maketitle

\begin{abstract}

If high mass di-gamma events observed at LEP are due to the
production of a di-gamma resonance via its leptonic coupling,
its consequences can be observed at TRISTAN.
We find that a predicted $Z$ decay branching rate is too small to
account for the observed events if the resonance spin is zero,
due to a strong cancellation in the decay amplitudes.
Such a cancellation is absent if the resonance has a spin two.
We study the consequences of a tensor production in the
processes $e^+e^- \to e^+e^-$, $\mu^+\mu^-$ and $\gamma\,\gamma$
at TRISTAN energies.
Complete helicity amplitudes with tensor boson exchange
contributions are given, and the signal can clearly be identified
from various distributions.
TRISTAN experiments are also sensitive to the virtual tensor boson
exchange effects, which reduce to the contact interaction terms in
the high mass limit.
\end{abstract}


\newpage
\section{Introduction}
\label{Introduction}

Following the observation by the L3 collaboration of a cluster of
events with a 60\,GeV di-gamma and a charged lepton pair\cite{L3},
similar events have been looked for by the other experimental
group at LEP\cite{kawamoto}.
Although the possibility exists that these events can be explained
away as a statistical fluctuation of normal radiative $Z$ decays,
it is worthwhile to study the consequences of a narrow di-gamma
resonance of mass about 60\,GeV\cite{barger,bando}.

When viewed as $Z$ decays into a di-gamma resonance,
the reported events\cite{L3,kawamoto} seem to indicate that it is
produced only in association with a charged lepton pair,
$e^+e^-$ or $\mu^+\mu^-$.
In particular, no apparent clustering of events with a di-gamma
invariant mass at around $60 {\mbox{GeV}}$ has been found in the
other channels, $\gamma \gamma \nu \overline{\nu}$ and
$\gamma \gamma q \overline{q}$.
This seems to exclude the possibility that a di-gamma resonance
is produced in association with a virtual $Z$ boson, as expected
for a Higgs-like boson\cite{barger}.

One possible explanation is that a di-gamma resonance is produced
in $Z$ decays via its direct coupling to charged leptons.
With a signifcant coupling to electrons, the resonance can be
produced at TRISTAN and affects the cross sections for the
processes $e^+e^- \to e^+e^-, \mu^+\mu^-,\gamma\,\gamma$
at $e^+e^-$ center of mass energy
$\sqrt{s} \simeq 60 \,{\mbox{GeV}}$.
Searches for such events have been recently carried out at TRISTAN
and the absence of any deviation from the standard model (SM)
expectation has been reported\cite{tristan}.
The resulting preliminary bounds on the resonance parameters have
been reported\cite{tristan} by assuming that it has spin
zero\cite{hollik}.

In this paper, we report that the production of a massive spinless
boson, whether it is a scalar or a pseudoscalar or their doublet,
via its coupling to charged leptons is strongly suppressed in $Z$
decays due to a cancellation of amplitudes for the almost
axial-vector-like $Z$ coupling to charged leptons.
The suppression is so strong that a 60\,GeV spinless boson
should necessarily have as large a width as its mass in order to
gain a $Z$ branching fraction of the order of $10^{-6}$.

We find on the other hand
that such a cancellation does not take place for the production
of a tensor (spin-2) boson in $Z$ decays and
that a narrow 60\,GeV tensor boson can be produced via its
leptonic coupling in $Z$ decays with a significant branching
fraction.
We therefore study the consequences of a massive tensor boson that
couples to charged leptons and two photons at $e^+e^-$ collider
energies.
Helicity amplitudes are given for the processes
$e^+e^- \to e^+e^-, \mu^+\mu^-$ and $\gamma\,\gamma$
such that the differential cross sections for arbitrarily
polarized $e^+e^-$ beams are obtained easily.
The angular distributions are found to be distinctive for
a spin-2 boson exchange.
In contrast to the spin-0 boson exchange cases\cite{hollik},
we find significant interference effects between the spin-2 boson
exchange amplitudes and the SM ones, due to the
chirality-conserving nature of the tensor boson-lepton coupling.
The interference effects allow low energy experiments to have a
good sensitivity for the exchange of a very heavy spin-2 boson.
We also compare the effects of a tensor boson exchange in the
heavy mass limit and those of the contact four-fermion\cite{eichten}
or two-fermion-two-gamma\cite{HH} interactions.

The paper is organized as follows.
In section \ref{Zdecays},
we study $Z$ decays into a massive boson and a charged lepton pair
for the spin-0 and spin-2 cases.
The boson-leptonic widths are expressed in terms of the $Z$ decay
branching fraction.
In section \ref{TensorExchange},
the consequences of a massive spin-2 boson exchange in
the processes $e^+e^- \to e^+e^-, \mu^+\mu^-$ and $\gamma\,\gamma$
are studied in detail.
The complete helicity amplitudes and the most general differential
cross sections for these processes are presented in this section.
In section \ref{LargeLimit},
we compare the effects of a tensor boson exchange
in the heavy mass limit with those of contact four-fermion and
two-fermion-two-gamma interactions.
Section \ref{Conclusion} summarizes our findings.

\section{$Z$ decays into a boson and a charged lepton pair}
\label{Zdecays}

In this section, we study the decay
   \begin{eqnarray*}
       Z \longrightarrow X \ell \, \overline{\ell}
   \end{eqnarray*}
via boson $X$ couplings to charged leptons
$\ell \, \overline{\ell}$ ($\ell = e$ or $\mu$).
The boson $X$ has a mass around 60\,GeV and decays subsequently
into two photons, and hence the $X$ spin cannot be one
due to Yang's theorem.
We study the two simplest cases spin zero and two.

\subsection{Spin-0 boson}

A doublet of spinless bosons, a scalar $\phi_{_S}$ and
a pseudoscalar $\phi_{_P}$, can have significant couplings
to light leptons without violating the leptonic
chiral invariance\cite{peccei}.
Although their couplings to two photons should necessarily violate
the chiral invariance\cite{delaguila}, stringent limits on
their leptonic couplings from the electron and muon
anomalous magnetic moment measurements can be avoided
in the one-loop order if the doublet masses are almost
degenerate\cite{leurer}.

We adopt the following effective Lagrangian\cite{hollik} for the
spinless boson doublet and the charged leptonic fields $\psi_\ell$
   \begin{equation}
      {\cal L} = \overline{\psi_\ell}
              \left( f_{_S}\phi_{_S} + if_{_P}\gamma_5\,\phi_{_P}
              \right)
              \psi_\ell    \, + \, \mbox{h.c.} \, ,
      \label{2.1}
   \end{equation}
for $\ell=e$ and $\mu$.
The leptonic chiral invariance is preserved in the above interaction
if the doublet masses are degenerate ($m_{_S} = m_{_P}$) and the
couplings are the same ($f_{_S}=f_{_P}$).
The $Z$ boson can decay into a boson and a charged lepton pair
via the two Feynman diagrams of Fig.\ref{fig.1}.
It is instructive to study the helicity amplitude of the decay in the
massless lepton limit.
If the final lepton is left handed, the $Z$ boson couples to
the right-handed lepton in the diagram (a) because of the helicity
flip nature of the spinless boson coupling (\ref{2.1}),
whereas in the diagram (b)
the $Z$ boson couples to the left-handed lepton.
If we denote the $Z$ boson lepton coupling as
   \begin{equation}
      {\cal L} = \sum_l \, \overline{\psi_\ell} \,\gamma^\mu
            \left( g_{_L} P_{\!_L} + g_{_R} P_{\!_R} \right)
            \psi_\ell Z_\mu \,,
      \label{2.2}
   \end{equation}
with the chiral projectors $P_{\!_L} = (1-\gamma_5)/2$ and
$P_{\!_R} = (1+\gamma_5)/2$,
the SM couplings are
   \begin{eqnarray}
       g_{_L} &=& g_{_Z} (-{1 \over 2}+{\sin^2 \theta_{_W}} )\,,
       \nonumber \\
       g_{_R} &=& g_{_Z} {\sin ^2 \theta_{_W}} \,,
       \label{2.3}
   \end{eqnarray}
with
$g_{_Z}=g/ \cos\theta_{_W}=e/ \sin\theta_{_W}\cos\theta_{_W}$.
For ${\sin ^2 \theta_{_W}} =0.23$, the two couplings have almost
the same magnitude and opposite signs.
We find that, for a massive spinless boson, the two amplitudes
tend to cancel because of this near cancellation of
the $Z$ boson leptonic couplings in eq.\,(\ref{2.3}).

More explicitly, the helicity amplitudes for the process
   \begin{equation}
       Z(q,\lambda) \,\longrightarrow \,
                \phi_{{_S},{_P}}(k) +\ell\,(p,\sigma)
              + \overline{\ell}\,({\overline{p}},\overline{\sigma})
       \label{2.4}
   \end{equation}
are non-vanishing for massless leptons only when the two lepton
helicities agree\,($\sigma =\overline{\sigma}$),
and they are given as
   \begin{eqnarray}
       M(\lambda,\sigma=L) &=& \hphantom{+}
         g_{_R} \overline{u}(p,L) \, (f_{_S} +i f_{_P} \gamma_5 )
         \, {1 \over {p \hspace{-1.8mm} \slash}
                   + {k \hspace{-2mm}   \slash} }\,
          \epsilon \hspace{-1.6mm} \slash (q,\lambda) \,
                       v({\overline{p}},L)
        \nonumber \\
       & & + \, g_{_L} \overline{u} (p,L) \,
          \epsilon \hspace{-1.6mm} \slash (q,\lambda)
         {-1 \over \overline{p} \hspace{-1.8mm} \slash
          + {k \hspace{-2mm} \slash} }
         (f_{_S} +i f_{_P} \gamma_5 ) \, v({\overline{p}},L)
       \label{2.5}
   \end{eqnarray}
for left-handed leptons ($\sigma = L$).
The amplitude for right-handed lepton ($\sigma=R$) production is
obtained from (\ref{2.5}) simply by replacing the index $L$ with
$R$.
The cancellation between the two terms of eq.(\ref{2.5}) can be
shown as follows.
We express the squared amplitudes summed over helicities as
   \begin{equation}
      \sum_{\lambda,\sigma} \left| M(\lambda,\sigma) \right|^2
            = (f_{_S}^2+f_{_P}^2)
              \left\{
                     (g_{_L} +g_{_R})^2 \,\Sigma_V (s_1,s_2)
                   + (g_{_L} -g_{_R})^2 \,\Sigma_A (s_1,s_2)
              \right\} \,,
      \label{2.6}
   \end{equation}
and find
   \begin{eqnarray}
      \Sigma_V &=& (s_1 s_2 -m^2 s )
                   \left({1 \over s_1^2 } +{ 1 \over s_2^2 } \right)
                    + {2 (s_2-m^2) (s_1-m^2) \over s_1 s_2} \,,
      \label{2.7}
      \\ && \nonumber \\
      \Sigma_A &=& (s_1-m^2)(s_2-m^2)
                     \left({1 \over s_1} - {1 \over s_2} \right)^2
                  +\left({2 \over s} - {m^2 \over s_1^2}
                                     - {m^2 \over s_2^2} \right)
                  (s -s_1 -s_2 +m^2) \,,
      \label{2.8}
   \end{eqnarray}
for the degenerate mass case ($m_{_S} = m_{_P} = m$) where
$s_1 = (k + p)^2$ and $s_2 =(k + {\overline{p}})^2$ are
Dalitz variables and $s=m_{_Z}^2$.
We show in Figs.\ref{fig.2}\,(a) and \ref{fig.2}\,(b)
the Dalitz distributions $\Sigma_V$ and $\Sigma_A$, respectively.
$\Sigma_V$ is the term which should be dominant if the $Z$
couplings to charged leptons were vector-like,
whereas $\Sigma_A$ dominates in the SM where the $Z$ couplings
to charged leptons are almost axial-vector-like.
The cancellation between the two amplitudes in $\Sigma_A$ is
clearly seen in Fig.\ref{fig.2}\,(b) along the symmetric line
$s_1=s_2$.

The $Z$ boson decay distribution is expressed as
   \begin{equation}
      d \Gamma = {1 \over 2 m_{_Z}} \cdot {1 \over 3} \cdot
                 \sum_{\lambda, \sigma}
                 \left| M(\lambda,\sigma) \right| ^2 d \Phi \,,
      \label{2.9}
   \end{equation}
with the phase space factor
   \begin{equation}
      d \Phi =  {1 \over 128 \pi^3 m_{_Z}^2}  \,ds_1 ds_2
      \label{2.10}
   \end{equation}
parametrized in terms of the Dalitz variables.
Upon integration over the phase space, we find for
$m_{_S}=m_{_P}=m$ :
   \begin{equation}
      \Gamma (Z \to \phi \, \ell \, \overline{\ell}\,)
      = {(f_{_S}^2 + f_{_P}^2 ) \,m_{_Z} \over 768\,\pi^3}
       \left[ (g_{_L} + g_{_R})^2 \,F_V(m^2 / m_{_Z}^2)
            + (g_{_L} - g_{_R})^2 \,F_A(m^2 / m_{_Z}^2) \right]
      \label{2.11}
   \end{equation}
where
   \begin{eqnarray}
      F_V(x) &=&  -2 +8 x - 6 x^2 -(1-2 x - 3x^2) \ln x
      \nonumber \\
             & &  -x^2 \left\{4\,{\rm Li}_2\,
                              \left({1\over 1+x}\right)
                            -{\pi^2 \over 3} + 2 \ln^2(1+x) -\ln^2 x
                        \right\} \,,
      \label{2.12}
      \\ \nonumber \\
      F_A(x) &=&   -{ 11\over 3} -5 x + 9 x^2 -{1 \over 3} x^3
                   -(1 +8 x +3 x^2) \ln x
      \nonumber \\
             & &   +x^2 \left\{ 4 {\rm Li}_2\,
                               \left({1 \over 1+x} \right)
                               -{\pi^2 \over 3}+2\ln^2(1+x) -\ln^2 x
                               \right\} \,.
      \label{2.13}
   \end{eqnarray}
Here ${\rm Li}_2$ is the dilogarithm
${\rm Li}_2(z) = -\displaystyle{\int_0^z {dt \over t} \ln (1-t)}$.
At $x= (59 \,{\mbox{GeV}} /91.187 \,{\mbox{GeV}} )^2 = 0.419$,
we find $F_V(x) = 0.0186$ and $F_A(x) = 0.000619$.
The boson leptonic width is
   \begin{equation}
      \Gamma (\phi \to \ell \,\overline{\ell})
       = {(f_{_S}^2 + f_{_P}^2 ) \, m \over 8\,\pi}
       \label{2.14}
   \end{equation}
for a degenerate doublet.
We therefore find from eqs.(\ref{2.11})--(\ref{2.14}) that
   \begin{equation}
      \Gamma (\phi \to \ell \,\overline{\ell})
           = (1.56 \times 10^{7} {\mbox{GeV}})
             \times
             {\rm Br}(Z \to \phi \,\ell \,\overline{\ell})
      \label{2.15}
   \end{equation}
for $m= \,59 \,\mbox{GeV}$, $\sin ^2 \theta_{_W} = 0.23$, and
$g_{_Z}^2=4\pi\alpha (m_{_Z}^2)/ \sin^2\theta_{_W}\cos^2\theta_{_W}$
with $ \alpha (m_{_Z}^2) = 1/128$.
If we require that the branching fraction
$\displaystyle{\sum_{\ell=e,\mu}}
{\rm Br}(Z \to \phi \,\ell \,\overline{\ell})$
to be larger than $10^{-6}$, then the leptonic width
$\displaystyle{\sum_{\ell=e,\mu}}
\Gamma (\phi \to \ell \,\overline{\ell})$ should be larger than
15\,GeV.
This is clearly incompatible with the narrow width assumption of
the di-gamma resonance.
It is clear that no improvement of the situation is possible by
introducing a splitting in the scalar and the pseudo-scalar masses.

\subsection{Spin-2 boson}

We therefore examine production of a spin-2 (tensor) boson
in $Z$ decays,
   \begin{equation}
       Z(q,\lambda) \,\longrightarrow\,
            \phi_{_T}(k,\tau) +\,\ell\,(p,\sigma)
          + \overline{\ell}\,({\overline{p}},\overline{\sigma}) \,,
       \label{2.16}
   \end{equation}
where the indices $\lambda$, $\tau$, $\sigma$ and $\overline{\sigma}$
denote helicities of the $Z$, the tensor $\phi_{_T}$, the lepton and
the antilepton, respectively.
The chirality conserving couplings of a tensor boson to a lepton
can be parametrized as
   \begin{equation}
      {\cal L} = {i \over 2} \phi_{_T}^{\mu\nu} \sum_{\alpha = L,R}
              f_\alpha^\ell
              \left[ \overline{\psi_\ell}_\alpha \gamma_\mu
                              (D_\nu {\psi_\ell}_\alpha)
                   -(\overline{D_\nu {\psi_\ell}_\alpha}) \gamma_\mu
                              {\psi_\ell}_\alpha \right] \,,
      \label{2.17}
   \end{equation}
where $D_\nu$ denotes the gauge-covariant derivative of the SM.
The two terms in the square bracket are shown explicitly so that
the Lagrangian is Hermitian even for an off-shell tensor boson,
whose effects will be studied in the following section.
It should be noted that the SM gauge invariance implies that the
tensor boson $\phi_{_T}$ is a singlet when it couples to
right-handed charged leptons ($f_{_R}^\ell \ne 0, f_{_L}^\ell=0$),
whereas it is a singlet or a member of a triplet when it couples
to left-handed charged leptons ($f_{_L}^\ell \ne 0,f_{_R}^\ell=0$).
The two couplings $f_{_L}^\ell$ and $f_{_R}^\ell$ can in general
co-exist.
However, non-observation of or $\gamma \gamma \nu \overline{\nu}$
type events disfavors scenarios where $\phi_{_T}$ couples
to left-handed leptons.
We will hence consider the case with $f_{_L}^\ell =0$,
although we give expressions for general couplings below.

The $Z$ boson decay (\ref{2.16}) proceeds via the Feynman diagrams
of Fig.\ref{fig.1}\,(b).
In contrast to the case of the spinless boson, the helicity
amplitudes are non-vanishing when $\sigma = -\overline{\sigma}$
in the massless lepton limit, and all the three diagrams of
Fig.\ref{fig.1}\,(b) give a contribution proportional to the
same $Z$ boson leptonic coupling for each helicity amplitude.
The amplitudes for the right-handed lepton ($\sigma=R$) can be
written as
   \begin{eqnarray}
      \lefteqn{ M(\lambda,\tau,\sigma = R) } \nonumber \\
        &=& \,g_{_R} f_{_R}^\ell \,\epsilon_{_T}^{\mu\nu}(k,\tau)^* \,
            \epsilon^\rho(q,\lambda) \nonumber \\
        & & \times \: \overline{u}(p,R)
        \left\{ {p_\nu \over s_1} \gamma_\mu
                    ( {p \hspace{-1.8mm} \slash}
                     +{k \hspace{-2mm}   \slash} )
                     \gamma_\rho
               +{\overline{p}_\nu \over s_2} \gamma_\rho
                    ( \overline{p} \hspace{-1.8mm} \slash
                       +{k \hspace{-2mm} \slash} ) \gamma_\mu
               -\gamma_\mu g_{\nu \rho}
        \right\} P_{\!_R} v({\overline{p}},L) \,,
        \label{2.18}
   \end{eqnarray}
where $s_1 = (p + k)^2$ and $s_2 = ({\overline{p}} + k)^2$.
Those for the left-handed lepton ($\sigma = L$) are obtained from
the above by the replacement $R \leftrightarrow L$.
The tensor wave functions $\epsilon_{_T}^{\mu \nu}$ should satisfy
the following conditions for each helicity $\tau$
   \begin{eqnarray}
      \epsilon_{_T}^{\mu \nu}(k,\tau)
        = \epsilon_{_T}^{\nu \mu}(k,\tau) \,,\qquad
      \label{2.19} \\
        k_\mu \, \epsilon_{_T}^{\mu \nu}(k,\tau) \,
      = k_\nu \, \epsilon_{_T}^{\mu \nu}(k,\tau) = 0 \,,
      \label{2.20} \\
      \epsilon_{_T}^{\mu \nu}(k,\tau) \, g_{\mu \nu} = 0 \,,
      \qquad \quad
      \label{2.21}
   \end{eqnarray}
and they can be normalized as
   \begin{equation}
      \epsilon_{_T}^{\mu \nu}(k,\lambda)^* \,
      {\epsilon_{_T}}_{\mu \nu}(k,\lambda')
       = \delta_{\lambda,\lambda'} \,.
      \label{2.22}
   \end{equation}
The completeness condition is\cite{veltman}
   \begin{equation}
      \sum_\tau {\epsilon_{_T}}^{\alpha \beta}(k,\tau) \,
                {\epsilon_{_T}}^{\mu \nu}(k,\tau)^*
      = {1 \over 2}\left( \kappa^{\mu \alpha}\kappa^{\nu \beta}
                        + \kappa^{\mu \beta} \kappa^{\nu \alpha}
                   \right)
       -{1 \over 3} \,\kappa^{\alpha \beta} \kappa^{\mu \nu}  \,,
      \label{2.23}
   \end{equation}
where
   \begin{equation}
      \kappa^{\mu \nu} = -g^{\mu \nu} + {k^\mu k^\nu \over m_{_T}^2}
      \label{2.24}
   \end{equation}
for the tensor mass $m_{_T}$.
It should be noted that tensor wave functions are obtained
as products of two vector wave functions.
The helicity zero ($\tau = 0$) component contains a product of two
longitudinary polarized vector wave functions and it behaves as
$E^4/m_{_T}^4$ at high energies.
Because of this high energy behavior, we cannot obtain sensible
cross sections for the tensor exchange processes in perturbation
theory at energies far above the mass of the tensor boson,
while we can regard the interaction (\ref{2.17}) as an effective
one valid at and below the tensor mass scale.

By squaring the amplitude and summing over helicities, we find
   \begin{equation}
      \sum_{\lambda, \tau, \sigma}
          \left| M(\lambda ,\tau, \sigma) \right|^2
      = m_{_Z}^2 \left[ (g_{_R} f_{_R}^\ell)^2
                       +(g_{_L} f_{_L}^\ell)^2 \right]
      \Sigma_T(s_1,s_2) \,,
      \label{2.25}
   \end{equation}
where we factor out $m_{_Z}^2$ to account for the mass inverse
dimension of the effective couplings $f_{_R}^\ell$ and
$f_{_L}^\ell$.
The function $\Sigma_T(s_1,s_2)$ is given as
   \begin{eqnarray}
      \Sigma_T(s_1,s_2) &=&
         \,-\, {10 \over 3} \,-\,{4s \over m^2}
         \,-\, {s_1+s_2 \over s}
                \left( 2 - {10s \over 3 m^2} \right)
         \,+\, {s_1 s_2 \over m^4}
                \left({2 \over 3}+{4 m^2 \over s} \right)
       \nonumber \\
       &&\,-\, { (2s+m^2)(3s+m^2) \over s}
               \left( {1\over s_1}+{1\over s_2} \right)
         \,+\, {s_1^2+s_2^2 \over 3\,m^4}\,
                \left( 1-{s_1 \over s}-{s_2 \over s}
                        +{m^2\over s} \right)
       \nonumber \\
       &&\,-\, {m^4\over 2}\,
               \left({1\over s_1^2} +{1\over s_2^2}\right)
         \,+\, \left(2+ {m^2 \over 2 s}\right)
               \left\{ {s_1 \over s_2}+{s_2 \over s_1}
                     +{2(s+m^2)^2 \over s_1 s_2} \right\} \,.
\end{eqnarray}
The Dalitz distribution $\Sigma_T (s_1,s_2)$ is shown in Fig.4.
By comparing with the distributions Fig.2\,(a) and (b)
for the spinless boson case,
we find that no subtle cancellation takes place in the $Z$ decay
into a tensor $\phi_{_T}$ and a charged lepton pair.

Upon integration over phase space, we find
   \begin{equation}
      \Gamma (Z \to \phi_{_T} \ell \, \overline{\ell})
          = { m_{_Z}^3 \over 1536 \pi^3 }
      \left[ (g_{_R} f_{_R}^\ell)^2 + (g_{_L} f_{_L}^\ell)^2 \right]
      F_T ( m_{_T}^2 / m_{_Z}^2 ) \,,
      \label{2.26}
   \end{equation}
where
   \begin{eqnarray}
    F_T(x) \; &=&  \; {1 \over 90} (1-x)
           \left(362\,x^2+2897\,x+2607-{113 \over x}+{7 \over x^2}
           \right)
            \nonumber \\
       & & +{1 \over 3} \ln x \, \left(9\,x^3+48\,x^2+97\,x+38\right)
            \nonumber \\
       & & -(x+1)^2\,(x+4) \Biggl\{ -\ln^2 x +\ln^2 (1+x)
                                   +2 \ln x \,\ln(1+x)
            \nonumber \\
       & & {\qquad \qquad \qquad \qquad \qquad \qquad \qquad }
                  +2 {\rm Li}_2\, (-x)\,+2 {\rm Li}_2\,
                     \left({1 \over x+1} \right)
                    \Biggr\} \,.
      \label{2.27}
   \end{eqnarray}
At $x=(59\,{\mbox{GeV}} /91.187\,{\mbox{GeV}})^2$,
we find $F_T(x) = 0.358$.
The leptonic width is found as
   \begin{equation}
      \Gamma (\phi_{_T} \to \ell \,\overline{\ell})
       = {(f_{_R}^\ell)^2 + (f_{_L}^\ell)^2 \over 320 \pi} \,
         m_{_T}^3 \,.
      \label{2.28}
   \end{equation}
We hence find
   \begin{equation}
      \Gamma (\phi_{_T} \to \ell \,\overline{\ell})
          = ( 3.05 \times 10^{4} {\mbox{GeV}} )
       \times { (f_{_R}^\ell)^2 + (f_{_L}^\ell)^2 \over
                (f_{_R}^\ell)^2
               +\displaystyle{\left({g_{_L} \over g_{_R}} \right)^2}
                          (f_{_L}^\ell)^2 }
       \times {\rm Br}(Z \to \phi_{_T} \ell \,\overline{\ell})
      \label{2.29}
   \end{equation}
for $m_{_T}=59\,{\mbox{GeV}}$ and ${\sin ^2 \theta_{_W}} = 0.23$.
Since the factor
$(g_{_L}/g_{_R})^2 = (({1\over 2} -{\sin^2 \theta_{_W}})
/ \sin^2 \theta_{_W})^2 \approx 1.38$
is almost unity, the required width is insensitive to the specific
coupling choice ($f_{_R}^\ell,f_{_L}^\ell$).
It is now possible for a narrow resonance of
$\displaystyle{\sum_{\ell=e,\mu}}
\Gamma(\phi_{_T}\to \ell \,\overline{\ell}) \sim 30\,{\mbox{MeV}}$
to be produced in $Z$ decays at a branching
fraction $10^{-6}$.

The factor of $\sim 500$ difference in the numerical factor for
the spinless boson production in eq.(\ref{2.15})
and that for the tensor boson production in eq.(\ref{2.29})
is semi-quantitatively understood as follows.
The cancellation in the dominant axial vector contribution $F_A(x)$
to the spinless boson production in eq.(\ref{2.11}) gives about a
factor of 30 suppression as compared to the vector part $F_V(x)$.
The tensor production factor $F_T(x)$ of eq.(\ref{2.26}) is even
larger than $F_V(x)$ partly because of the spin counting factor
$2 \cdot 2 +1 =5$.
Finally, the higher dimensionality of the effective tensor coupling
in the Lagrangian (\ref{2.17}) leads to the counting factor of
$(m_{_Z} / m_{_X})^2 \approx 2$.
The factors add up to about 300, almost explaining the drastic
change in the $Z$ decay branching fractions between the spin zero
and two cases.

\section{Tensor exchange in \lowercase{$e^+e^-$} colliders}
\label{TensorExchange}

Effects of a spinless boson exchange in $e^+e^-$ collisions have
been rather thoroughly studied in refs.\cite{hollik,hikasa}.
In this section, we study in detail the consequences of a tensor
boson exchange in the processes $e^+e^- \to e^+e^-,\,\mu^+\mu^-$
and $\gamma\,\gamma$.

Here we assume that the $\phi_{_T}$ couplings to the leptons and
that with two photons can be parametrized by the Lagrangian
   \begin{equation}
      {\cal L} = \phi_{_T}^{\mu\nu}
      \left\{{i\over 2} \,\sum_\ell \sum_{\alpha = L,R}f_\alpha^\ell
         \left[ \overline{\psi_\ell}_\alpha \gamma_\mu
                         (D_\nu {\psi_\ell}_\alpha)
              -(\overline{D_\nu {\psi_\ell}_\alpha}) \gamma_\mu
                         {\psi_\ell}_\alpha \right]
         +e^2 h \,F_{\mu \lambda} {F^\lambda}_\nu
      \right\} \,,
      \label{3.1}
   \end{equation}
where the first term in the curly brackets is the
$\phi_{_T} \ell \,\overline{\ell}$ coupling of eq.(\ref{2.17}),
and $F_{\mu\lambda}=\partial_\mu A_\lambda -\partial_\lambda A_\mu$
is the gauge invariant field strength of the electromagnetic field.
The leptonic and the photonic widths are
   \begin{eqnarray}
      \Gamma (\phi_{_T} \to \ell \,\overline{\ell})
       &=& {(f_{_R}^\ell)^2 + (f_{_L}^\ell)^2 \over 320 \pi}
           \,m_{_T}^3 \,,
      \label{3.2}\\
      \Gamma (\phi_{_T} \to \gamma \,\gamma)
       &=& {\pi \alpha^2 h^2 \over 5} \,m_{_T}^3 \,,
      \label{3.3}
   \end{eqnarray}
and the total $\phi_{_T}$ width may be written as
   \begin{equation}
      \Gamma_{\!_T} = \sum_\ell
         \Gamma (\phi_{_T} \to \ell \,\overline{\ell})
       + \Gamma (\phi_{_T} \to \gamma\,\gamma) \,,
      \label{3.4}
   \end{equation}
if other decay modes of $\phi_{_T}$ are negligible.
Summation over lepton flavors should contain $\ell =e$ and $\mu$,
but it may or may not include the $\ell=\tau$ case.

As for the $\phi_{_T}$ propagator, we take the {\it unitary gauge}
form
   \begin{equation}
      D^{\mu\nu\alpha\beta}(k^2,m_{_T}^2)
      = {i \displaystyle{\sum_\tau}
             \epsilon_{_T}^{\mu\nu}(k,\tau)^*
             \epsilon_{_T}^{\alpha\beta}(k,\tau) \over
                       k^2 -m_{_T}^2 +i m_{_T} \Gamma_{\!_T} }
      \label{3.5}
   \end{equation}
with the spin summation factor given by eqs.(\ref{2.23}) and
(\ref{2.24}).
This form behaves as $E^2/m_{_T}^4$ at high energies
($|k^2| \gg m_{_T}^2$)
and cannot be used at energies significantly above the mass shell.
It may be regarded as an effective propagator near and below the
tensor mass scale, $|k^2| \lesssim m_{_T}^2$.

\subsection{$e^+e^- \to e^+e^-$}

We denote the helicity amplitude for the process
   \begin{equation}
      e^-(p,\sigma)  + e^+({\overline{p}},\overline{\sigma})
      \,\longrightarrow\,
      e^-(k,\lambda) + e^+({\overline{k}},\overline{\lambda})
      \label{3.6}
   \end{equation}
as
   \begin{equation}
      M_{\,\sigma,\overline{\sigma}}^{\lambda, \overline{\lambda}}
      \label{3.7}
   \end{equation}
where the indicies $\sigma$, $\overline{\sigma}$, $\lambda$,
$\overline{\lambda}$ are the helicities in units of
$\hbar/2$\cite{HZ1}.
There are six non-vanishing helicity amplitudes
   \begin{equation}
      M_{\,-+}^{-+},\, M_{\,-+}^{+-},\, M_{\,+-}^{+-},\,
      M_{\,+-}^{-+},\, M_{\,++}^{++},\, M_{\,--}^{--}\,,
      \label{3.8}
   \end{equation}
in the massless electron limit.
The most general differential cross section is expressed in terms
of these helicity amplitudes as\cite{HZ2}
   \begin{eqnarray}
      {d\sigma \over d\cos\theta \,d\phi} &=& {1 \over 256 \pi^2 s}
      \,\Biggl\{
               \left(1-P_-^L \right) \left(1+P_+^L \right)
               \left(  \left| M_{\,-+}^{-+} \right|^2
                     +\left| M_{\,-+}^{+-} \right|^2 \right)
                \nonumber \\
            & & \qquad \quad
              +\left(1+P_-^L \right) \left(1-P_+^L \right)
               \left( \left| M_{\,+-}^{+-} \right|^2
                    +\left| M_{\,+-}^{-+} \right|^2 \right)
                 \nonumber \\
            & & \qquad \quad
              +\left(1+P_-^L \right) \left(1+P_+^L \right)
               \left| M_{\,++}^{++} \right|^2
                 \nonumber \\
            & & \qquad \quad
              +\left(1-P_-^L \right) \left(1-P_+^L \right)
               \left| M_{\,--}^{--} \right|^2
                \nonumber \\
            & & \qquad \quad
              +2 P_-^T P_+^T \cos 2 \phi
               \; {\rm Re}
               \left[
                   \left( M_{\,-+}^{-+} \right)
                   \left( M_{\,+-}^{-+} \right)^*
                  +\left( M_{\,-+}^{+-} \right)
                   \left( M_{\,+-}^{+-} \right)^*
               \right]
                 \nonumber \\
            & & \qquad \quad
              +2 P_-^T P_+^T \sin 2 \phi
               \; {\rm Im}
               \left[
                   \left( M_{\,-+}^{-+} \right)
                   \left( M_{\,+-}^{-+} \right)^*
                  +\left( M_{\,-+}^{+-} \right)
                   \left( M_{\,+-}^{+-} \right)^*
               \right]
        \Biggr\} \,,
        \nonumber \\
      \label{3.9}
   \end{eqnarray}
where $\theta$ is the polar angle of the final $e^-$ about the
$e^-$ beam axis, $\phi$ is the azimuthal angle about the $e^-$ beam
axis measured from the $e^-$ natural polarization axis in the
storage ring, $P_{\pm}^L$ are the $e^\pm$ beam longitudinal polarization,
and $P_{\pm}^T$ are the $e^\pm$ beam transverse polarization.
We can set $P_+^L = P_-^L = 0$ and $P_+^T = P_-^T = P_T$ at TRISTAN.

The six non-vanishing helicity amplitudes are found as follows :
   \begin{eqnarray}
      M_{\,\sigma,-\sigma}^{\sigma,-\sigma} &=& -s (1+ \cos \theta)
      \Biggl\{ \sum_V \left( g_\sigma^{V\ell\ell} \right)^2
                     \left[D_{_V}(s)+D_{_V}(t) \right]
               \nonumber \\ & & {\qquad \qquad \qquad \qquad}
             +{\left( f_\sigma^e \right)^2 s \over 8}
               \left[ (2 \cos \theta -1) D_{_T}(s)
                     +{7+\cos \theta \over 2} D_{_T}(t) \right]
      \Biggr\} \,,
      \label{3.10}
      \\
      M_{\,\sigma, -\sigma}^{-\sigma, \sigma} &=& s (1-\cos\theta)
      \Biggl\{ \sum_V \, g_{_L}^{V\ell\ell}\,
                         g_{_R}^{V\ell\ell}\, D_{_V}(s)
             + {f_{_L}^e \, f_{_R}^e \, s \over 8}
                (2 \cos \theta +1) D_{_T}(s)
      \Biggr\} \,,
      \label{3.11}
      \\
      M_{\,\sigma,  \sigma}^{\sigma,  \sigma} &=&   -2s
      \Biggl\{ \sum_V \, g_{_L}^{V\ell\ell}\, g_{_R}^{V\ell\ell}\,
               D_{_V}(t)
             + { \,f_{_L}^e \, f_{_R}^e \, s \over 16}
              (5 +3 \cos \theta ) D_{_T}(t) \,
      \Biggr\} \,,
      \label{3.12}
   \end{eqnarray}
where $t=-s(1-\cos \theta)/2$, and the propagator factors are
   \begin{equation}
      D_{_B}(q^2) = {1 \over q^2-m_{_B}^2
                                +i m_{_B}\Gamma_{\!_B}\theta (q^2)}
      \label{3.13}
   \end{equation}
for $B=V$($\gamma$ or $Z$) and $T$.
We adopt a chirality index convention $\sigma=-$ for $\sigma=L$,
$\sigma=+$ for $\sigma=R$\cite{HZ1}.
The SM couplings are
   \begin{eqnarray}
      {g_{_L}^{\gamma \ell \ell}} &=& {g_{_R}^{\gamma \ell \ell}}
                                   =-e \,,
      \nonumber \\
      {g_{_L}^{Z \ell \ell}} &=& g_{_Z}
       \left( -{1\over 2} +{\sin ^2 \theta_{_W}} \right) \,,
      \label{3.14}\\
      {g_{_R}^{Z \ell \ell}} &=& g_{_Z} {\sin ^2 \theta_{_W}} \,,
      \nonumber
   \end{eqnarray}
with
$g_{_Z}=g/ \cos\theta_{_W} = e/ \sin \theta_{_W} \cos \theta_{_W}$.

The most general differential cross section is easily obtained
by inserting the amplitudes (\ref{3.10})--(\ref{3.12})
into (\ref{3.9}).
We show in Fig.~5 the energy dependence of the cross section for the
process $e^+e^- \to e^+e^-$ at $\cos\theta=0$, 0.4 and 0.8, together
with the SM predictions.
We chose the parameter values $f_{_R}^\ell=0.0086$, $f_{_L}^\ell=0$
and $\Gamma_T=100$\,MeV, which give $\sum_{\ell=e,\mu}{\rm Br}
(Z\to\phi_T\ell\overline{\ell})\sim 10^{-6}$ and ${\rm Br}(\phi_T \to
\gamma\gamma)\sim 0.7$.
At large scattering angles, slightly destructive interference is
observed below the resonance peak while above the peak a rather
large constructive interference effect is expected.
At smaller scattering angles where the $t$-channel $\gamma$ exchange
amplitude dominates the SM amplitude, the interference below the
resonance is found to be constructive.
Despite the narrow width of the resonance, its effect is found to be
significant in the wide range of the colliding beam energy due to the
interference effect.
TRISTAN experiments should hence be sensitive to the tensor boson
effects even when its width is much narrower than posturated here.

We find that the peak cross section is
   \begin{eqnarray}
      \sigma(e^+e^- \to \phi_{_T} \to e^+ e^-)_{s=m_{_T}^2}
        \:&=&\: {4 \pi \over m_{_T}^2} \,(2 \cdot 2+1) \,
              {\Gamma (\phi_{_T} \to e^+e^-)^2 \over \Gamma_{\!_T}^2}
      \label{3.15}
      \\
        \:&=&\: {\left[(f_{_R}^e)^2 +(f_{_L}^e)^2 \right]^2
             \over 5120 \pi} \,
            { m_{_T}^4 \over \Gamma_{\!_T}^2} \,.
      \label{3.16}
   \end{eqnarray}
The polar angle distribution at the peak is
   \begin{eqnarray}
       \lefteqn{ \left( {d\sigma \over d\cos \theta }
                 \right)_{s=m_{_T}^2}
        = \; \sigma (e^+e^- \to \phi_{_T} \to e^+e^-)_{s=m_{_T}^2}   }
        \nonumber \\
        &&  \times \left\{ \,{5 \over 8} (1+\cos \theta)^2
                            (1-2\cos \theta)^2
                     +{5 (f_{_R}^e)^2 (f_{_L}^e)^2 \over
                        \left[ (f_{_R}^e)^2 +(f_{_L}^e)^2
                        \right]^2 }
                     \cos \theta (1-2\cos^2 \theta)  \,\right\} \,.
      \label{3.17}
   \end{eqnarray}
%

\subsection{$e^+e^- \to \mu^+ \mu^-, \,\tau^+ \tau^-$}

For the process
   \begin{equation}
      e^-(p,\sigma) + e^+({\overline{p}},\overline{\sigma})
      \,\longrightarrow\,
           \ell\,(k,\lambda) + \,
           \overline{\ell}\,({\overline{k}}, \overline{\lambda})
      \label{3.18}
   \end{equation}
with $\ell=\mu$ or $\tau$, only the first four helicity
amplitudes of (\ref{3.8}) are non-vanishing in the massless lepton
limit.
They are obtained from eqs.(\ref{3.10}) and (\ref{3.11}) simply by
replacing the leptonic coupling factors and by dropping the
$t$-channel exchange contribution :
   \begin{eqnarray}
      M_{\,\sigma, -\sigma}^{\sigma, -\sigma}
      &=& -s\,(1+\cos\theta)
      \left\{\,\sum_V \, g_\sigma^{Vee}\, g_\sigma^{V\ell\ell} \,
                    D_{_V}(s)
             +{1\over 8} \,f_\sigma^e f_\sigma^\ell s \,
                         (2 \cos \theta -1 )  \, D_{_T}(s) \,
      \right\} \,,
      \label{3.19}
      \\
      M_{\,\sigma, -\sigma}^{-\sigma, \sigma} &=& \hphantom{-}
      s\,(1- \cos\theta)
      \left\{\, \sum_V \, g_\sigma^{Vee} \, g_{-\sigma}^{V\ell\ell}
                     \, D_{_V}(s)
            +{1 \over 8} \,f_\sigma^e \,f_{-\sigma}^\ell
                         s (2 \cos \theta +1)  \, D_{_T}(s) \,
      \right\} \,.
      \label{3.20}
   \end{eqnarray}
The most general differential cross section is then obtained by
inserting the above helicity amplitudes into the formula (\ref{3.9}).
The peak cross section is
   \begin{equation}
      \sigma (e^+e^- \to \phi_{_T} \to \ell^+ \ell^-)_{s=m_{_T}^2}
       \:=\: {4 \pi \over m_{_T}^2} \,(2 \cdot 2+1) \,
          { \Gamma(\phi_{_T} \to e^+e^-) \,
            \Gamma(\phi_{_T} \to \ell^+\ell^-)
           \over \Gamma_{\!_T}^2 } \,,
      \label{3.21}
   \end{equation}
and the polar angle distribution at the peak is
   \begin{eqnarray}
     \left( {d\sigma \over d\cos \theta } \right)_{s=m_{_T}^2}
       &=&\: \sigma (e^+e^-\to \phi_{_T} \to\ell^+\ell^-)_{s=m_{_T}^2}
       \nonumber \\
       & & \times \:
           \Biggr\{\, {5 \over 8}(1+\cos \theta)^2 (1-2\cos\theta)^2
       \nonumber \\
       & &  \qquad
          +{5 \, f_{_R}^e f_{_L}^e f_{_R}^\ell f_{_L}^\ell
            \over  \left[ (f_{_R}^e)^2    +(f_{_L}^e)^2     \right]
                   \left[ (f_{_R}^\ell)^2 +(f_{_L}^\ell)^2  \right] }
                     \cos\theta\, (1-2\cos^2 \theta)  \Biggr\}
      \,. \quad
      \label{3.22}
   \end{eqnarray}

We show in Fig.~6 the energy dependence of the differential cross
section for the process $e^+ e^- \to \mu^+ \mu^-$ at
$\cos\theta =0$, together with the SM prediction depicted
by the dashed line.
The tensor couplings to left-handed leptons have been set to zero,
$f_{_L}^e=f_{_L}^\mu=0$,as before.
Significant interference effect is again expected for the posturated
resonance parameters
($m_{_T} =59$GeV and $\Gamma(\phi_{_T} \to e^+e^-)=15$MeV).
The interference below the resonance peak is constructive
(distractive) if the tensor couplings to electrons\,($f_{_R}^e$)
and those to muons\,($f_{_R}^\mu$) have the same (opposite) sign.
It is clear from the figure that TRISTAN experiments should be
sensitive to a tensor resonance with much smaller leptonic width.

\subsection{$e^+e^- \to \gamma\,\gamma$}

We denote the helicity amplitude for the process
   \begin{equation}
      e^-(p,\sigma) +e^+({\overline{p}},\overline{\sigma})
      \,\longrightarrow\,
      \gamma\,(k_1,\lambda_1)  + \, \gamma\,(k_2,\lambda_2)
      \label{3.23}
   \end{equation}
as
   \begin{equation}
      M_{\,\sigma \; , \; \overline{\sigma}}^{\lambda_1, \lambda_2}
      \label{3.24}
   \end{equation}
where $\lambda_i = \pm$ denotes the $\gamma$ helicities in units
of $\hbar$.
We find in the convention of ref.\cite{HZ1} that only the following
4 helicity amplitudes ($\sigma =\pm$ and $\lambda = \pm$)
   \begin{equation}
      M_{\,\sigma,-\sigma}^{\lambda,-\lambda}
      \,=\, 2 \,\sigma \lambda \,e^2 {1 + \sigma\lambda\cos\theta
                              \over \sin \theta}
        \left[ 1 + {f_\sigma^e h \,s^2 \sin^2 \theta \over
                    8 \,(m_{_T}^2 -s- im_{_T}\Gamma_{\!_T}) }  \right]
      \label{3.25}
   \end{equation}
are non-vanishing in the massless electron limit.
It is worth noting that both the SM and the tensor boson exchange
contributions satisfy the {\it helicity conservation} condition,
$\lambda_2=-\lambda_1$, for the two out-going photons.
This leads to significant interference effects between the two
contributions.

The general differential cross section is again obtained by inserting
the helicity amplitudes (\ref{3.25}) into the generic formula
(\ref{3.9}).
For $P_+^L = P_-^L = 0$ and $P_+^T = P_-^T = P_T$,
we find more explicitly
   \begin{eqnarray}
      {d\sigma \over d\cos\theta \,d\phi} \;
        = \; {\alpha^2 \over s} \Biggl\{
       & & {1+\cos^2 \theta \over \sin^2 \theta}
       \nonumber \\
       &+& { \left( f_{_R}^e+f_{_L}^e \right) h s^2 (m_{_T}^2 -s)
            \over
            8 \left[ (m_{_T}^2 -s)^2 +(m_{_T}\Gamma_{\!_T})^2 \right] }
            \,(1+\cos^2 \theta)
       \nonumber \\
       &+& { \left[ \left( f_{_R}^e \right)^2
                   +\left( f_{_L}^e \right)^2 \right]
                 h^2 s^4
            \over
            128 \left[ (m_{_T}^2 -s)^2 +(m_{_T}\Gamma_{\!_T})^2 \right] }
            \,\sin^2 \theta \,(1+\cos^2 \theta)
       \nonumber \\
       &-& P_{_T}^2 \Biggl[ 1+
            { \left( f_{_R}^e+f_{_L}^e \right) h s^2 (m_{_T}^2 -s)
             \over
            8 \left[ (m_{_T}^2 -s)^2 +(m_{_T}\Gamma_{_T})^2 \right] }
              \,\sin^2 \theta
       \nonumber \\
       & & \qquad +
            { f_{_R}^e f_{_L}^e h^2 s^4
            \over
            64 \left[ (m_{_T}^2 -s)^2 +(m_{_T}\Gamma_{\!_T})^2\right]}
              \,\sin^4 \theta    \Biggr] \cos 2 \phi
       \nonumber \\
       &-& P_{_T}^2
            { \left(-f_{_R}^e+f_{_L}^e \right) hs^2m_{_T}\Gamma_{\!_T}
             \over
             8 \left[ (m_{_T}^2 -s)^2 +(m_{_T}\Gamma_{\!_T})^2\right]}
              \,\sin^2 \theta \,\sin 2 \phi  \quad \Biggr\} \,.
      \label{3.26}
   \end{eqnarray}
The peak cross section is
   \begin{equation}
      \sigma (e^+e^- \to \gamma \,\gamma)_{s=m_{_T}^2}
        \,=\, {4 \pi \over m_{_T}^2} (2 \cdot 2+1)
              {\Gamma(\phi_{_T} \to e^+e^-) \,
               \Gamma(\phi_{_T} \to \gamma \,\gamma)
               \over \Gamma_{\!_T}^2} \,,
      \label{3.27}
   \end{equation}
and the angular distribution at the peak is
   \begin{eqnarray}
       \lefteqn{ \left( {d\sigma \over d\cos\theta \,d\phi}
                 \right)_{s=m_{_T}^2}
       = \, \sigma (e^+e^- \to \gamma \,\gamma)_{s=m_{_T}^2} \cdot
              {5 \over 8 \pi} \sin^2 \theta  }\nonumber \\
       & & \times
        \left\{ 1+\cos^2 \theta
                +P_{_T} \left[{2 f_{_R}^e f_{_L}^e \sin^2 \theta
                               \over
                               (f_{_R}^e)^2 +(f_{_L}^e)^2 }
                               \cos 2\phi
                             \,+\,
                              {16 (-f_{_R}^e +f_{_L}^e)\Gamma_{\!_T}
                               \over
                               \left[ (f_{_R}^e)^2 +(f_{_L}^e)^2
                               \right]
                                h m_{_T}^3 }
                               \sin 2\phi \right]  \right\} \,.
      \label{3.28}
       \nonumber \\
   \end{eqnarray}

Shown in Fig.~7 is the energy dependence of the expected
differential cross section at $\cos\theta=0$ for $m_{_T}=59$GeV,
$\Gamma(\phi_{_T} \to e^+e^-)=15$MeV and
$\Gamma(\phi_{_T} \to \gamma \gamma)=70$MeV
($\Gamma_{\!_T}=100$MeV ).
The SM prediction is given by the dashed line.  Huge interference
effect is expected for the above resonance parameters.
The effect below the resonance is constructive
if the couplings $f_R^e$ and $h$ have the common sign,
wehreas it is distractive otherwise.
We may conclude from Figs.~5, 6 and 7 that the tnesor resonance
effect can be most clearly observed in the
$e^+ e^- \to \gamma\gamma$ channel.

\section{Large tensor mass limits}
\label{LargeLimit}

We found in the previous section that the effects of a tensor boson
exchange can be observbed at energies far away from the resonance
peak due to the interference between the tensor exchange amplitudes
and the SM ones.
This is in constrast to the spinless boson exchange case [6]
where a significant interference effect is expected only in the
$e^+e^- \to e^+e^-$ channel.
The difference arises because the tensor boson coupling to leptons
can be helicity conserving just like the SM gauge boson couplings
while the spinless boson coupling should neccessarily flip the
lepton helicities.
Therefore, $e^+ e^-$ collider experiments can have sensitivity to
tensor boson exchange effects even when its mass is far above the
colliding beam energy.

In the large $m_{_T}$ limit ($s/m_{_T}^2 \ll 1$), the $\phi_{_T}$
exchange contributions to the processes
$e^+e^- \to \ell \,\overline{\ell}$ are related to those of the
standard contact $ee\ell\ell$ interaction\cite{eichten} at
fixed energy and scattering angle.
Note that we cannot simply substitute the limits on the combination
$ f_\sigma^e f_\sigma^\ell / m_{_T}^2 $ by those of the corresponding
contact terms independent of energies and angles,
because the $\phi_{_T}$ exchange gives effective 4-fermion operators
of spin-2 and dimension-8 whereas the standard contact terms have
spin-1 and dimension-6.

On the other hand, the dimension-6 contact $ee\gamma\gamma$
interaction terms as listed by ref.\cite{buchmuller} give vanishing
matrix elements for the process $e^+e^- \to \gamma \,\gamma$.
It has been found\cite{HH} that the lowest dimensional local
operator that is chiral invariant and has non-vanishing
$e^+e^- \to \gamma \,\gamma$ matrix elements has dimension-8 and
spin-2.
Therefore, in the large $m_{_T}$ limit, the $\phi_{_T}$ exchange
contribution reduces to the dimesion-8 contact $ee\gamma\gamma$
term in the process $e^+e^- \to \gamma \,\gamma$.

Let us recall the standard dimension-6 contact interaction for
leptons\cite{eichten,pdg}
   \begin{equation}
      {\cal L}={g^2 \over \Lambda^2}
      \sum_{(\ell,\ell')} \, \sum_{\alpha=L,R} \, \sum_{\beta=L,R}
      \eta_{\alpha\beta}^{\ell\ell'} \,S^{\ell\ell'}\,
      \overline{\psi_\ell}_\alpha \gamma^\mu {\psi_\ell}_\alpha
      \overline{\psi_\ell'}_\beta \gamma_\mu {\psi_\ell'}_\beta
      \label{4.1}
   \end{equation}
where $g^2=4\pi$ by convention,
$|\eta_{\alpha\beta}^{\ell\ell'}| = 1$ or $0$
for special cases, and
   \begin{equation}
      S^{\ell \ell'}
      = \left\{ \begin{array}{ccc}
                       1      \quad &\mbox{for} &\quad \ell \ne \ell'
                               \\[-1mm]
                  {1 \over 2} \quad &\mbox{for} &\quad \ell  =  \ell'
                \end{array}
        \right.
      \label{4.2}
   \end{equation}
is the statistical factor.

We find that the helicity amplitudes for the process
$e^+e^- \to \ell\,\overline{\ell}$ (\ref{3.19})--(\ref{3.20})
reproduce those with the above contact interaction terms by
the following substitutions :
   \begin{eqnarray}
      {f_\sigma^e f_\sigma^\ell \, s \over 8} (2\cos\theta-1)
      D_{_T}(s)
      \:&\longrightarrow&\: {g^2 \eta_{\sigma\sigma}^{e\ell}
                             \over \Lambda^2}  \,,
      \label{4.3}
      \\
      {f_\sigma^e f_{-\sigma}^\ell \, s \over 8} (2\cos\theta+1)
      D_{_T}(s)
      \:&\longrightarrow&\: {g^2 \eta_{\sigma,-\sigma}^{e\ell}
                            \over \Lambda^2}  \,.
      \label{4.4}
   \end{eqnarray}
In the large mass limit, the following relationships hold
   \begin{eqnarray}
      {f_\sigma^e f_\sigma^\ell s \over 8 m_{_T}^2} (1-2 \cos\theta)
      &=& {g^2 \eta_{\sigma\sigma}^{e\ell} \over \Lambda^2}  \,,
      \label{4.5}
      \\
      -{f_\sigma^e f_{-\sigma}^\ell s \over 8 m_{_T}^2}
       (1+2 \cos \theta)
      &=& {g^2 \eta_{\sigma,-\sigma}^{e\ell} \over \Lambda^2} \,.
      \label{4.6}
   \end{eqnarray}
The helicity amplitudes for the process $e^+e^- \to  e^+e^-$
(\ref{3.10})--(\ref{3.12}) reduce to those with the contact $eeee$
term via the substitutions
   \begin{eqnarray}
      &&{ \left(f_\sigma^e \right)^2 s \over 8 }
        \left[ (2\cos\theta -1) D_{_T}(s)
              +{1\over 2} (7+\cos\theta) D_{_T}(t) \right]
        \; \longrightarrow \; {2 g^2\eta_{\sigma,\sigma}^{ee}
                               \over \Lambda^2} \,,
      \label{4.7}
       \\
      &&{ f_{_L}^e f_{_R}^e \, s \over 8 }
          (2\cos\theta +1) D_{_T}(s)
        \; \longrightarrow \;  { g^2 \eta_{{_L}{_R}}^{ee}
                                 \over \Lambda^2} \,,
      \label{4.8}
       \\
      &&{ f_{_L}^e f_{_R}^e \, s \over 16 }
          (5 +3\cos\theta ) D_{_T}(t)
         \; \longrightarrow \;  { g^2 \eta_{{_L}{_R}}^{ee}
                                 \over \Lambda^2} \,.
      \label{4.9}
   \end{eqnarray}
In the $m_{_T}^2 \gg s$ limit, the substitution rule (\ref{4.7})
gives the identification
   \begin{equation}
      - {5 (f_\sigma^e)^2 \, s \over 16 m_{_T}^2} (1+\cos \theta)
      = {2g^2\eta_{\sigma\sigma}^{ee} \over \Lambda^2} \,.
      \label{4.10}
   \end{equation}
However, the rules (\ref{4.8}) and (\ref{4.9}) do not allow us to
relate the parameters $\eta_{{_L}{_R}}^{ee}$ consistently with the
corresponding terms of the $\phi_{_T}$ exchange parameters even in
the large $m_{_T}$ limit at fixed energy and scattering angle.

The lowest dimension chirality conserving $ee\gamma\gamma$ operator
which gives non-vanishing $e^+e^- \to \gamma\,\gamma$ matrix
elements is found to be\cite{HH}
   \begin{equation}
      {\cal L} = {2\,i\,e^2 \over \Lambda^4}\,
                 F^{\mu\sigma} {F_\sigma}^\nu \sum_{\alpha=L,R}
                \eta_\alpha \overline{\psi}_\alpha \gamma_\mu
                \partial_\nu \psi_\alpha \,.
      \label{4.11}
   \end{equation}
The helicity amplitudes with the above contact terms are
   \begin{equation}
      M_{\,\sigma,-\sigma}^{\lambda,-\lambda}
      \,=\, 2 \,\sigma\lambda \,e^2 \,
           {1+\sigma\lambda\cos\theta \over \sin\theta}
          \left[1+ {\eta_\sigma\,s^2\sin^2\theta \over 4\Lambda^4}
          \right] \,.
      \label{4.12}
   \end{equation}
By comparing the above amplitudes and the $\phi_{_T}$ exchange
ones (\ref{3.25}), we find the substitution rule
   \begin{equation}
      {f_\sigma^e h \over 8 (m_{_T}^2 -im_{_T}\Gamma_{\!_T} -s )}
      \; \longrightarrow \;{\eta_\sigma \over 4 \Lambda^4} \,.
      \label{4.13}
   \end{equation}
In the large $m_{_T}$ limit ($m_{_T}^2 \gg s$),
we can make the following identification
   \begin{equation}
      {f_\sigma^e h \over 8 m_{_T}^2}
       = {\eta_\sigma \over 4\Lambda^4} \,,
      \label{4.14}
   \end{equation}
which is valid at all energies and angles.

The relationships between the tensor exchange amplitudes and
the contact $ee\ell\ell$ and $ee\gamma\gamma$ interactions may
be useful to estimate bounds on the $\phi_{_T}$ parameters in the
$m_{_T}^2 \gg s$
limit from the existing limit on the contact interactions.

\section{Conclusions}
\label{Conclusion}

Assuming that high mass di-gamma events observed at LEP
are caused by a di-gamma resonance which is produced via its
leptonic couplings,
we explored its consequences at TRISTAN. We examined both spin-0
and spin-2 cases as the simplest possibilities.

We showed that, in order to account for the branching fraction
${\rm Br}(Z\rightarrow \phi_{{_S},{_P}}\ell\bar\ell)$
of the order of $10^{-6}$,
the couplings of a spin-0 boson to leptons have to be so large
that the boson width should be comparable to its mass of
about 60\,GeV.
This is because of a strong cancellation in the decay amplitudes
for the almost axial vector $Z\ell\bar\ell$ coupling of the SM,
which causes a destructive interference between the two
Feynman diagrams of Fig.\ref{fig.1}.
As a result, the spin-0 case is excluded whether it is a scalar
($\phi_{_S}$), or a pseudo scalar ($\phi_{_P}$), or their doublet.

For the case of a spin-2 boson ($\phi_{_T}$),
no such cancellation arises
because it has chirality-conserving couplings to a lepton.
We found that a narrow resonance of about 30 \,MeV leptonic width
can be produced in $Z$ decays at the required branching
fraction level.
A higher spin resonance may improve the situation further
by means of spin and dimensional counting.

If there is such a di-gamma tensor resonance around 60\,GeV,
we can observe its effect at TRISTAN.
We gave complete helicity amplitudes including tensor boson
exchange for the processes $e^+e^- \to e^+e^-, \,\mu^+\mu^-$
and $\gamma\,\gamma$.
Differential cross sections for arbitrarily polarized beams
are easily obtained from the amplitudes.
The total cross sections and angular distributions behave
quite differently from the spin-0 resonance cases\cite{hollik}.
The resonance spin can hence be determined from these distributions.

We further noted  that the tensor resonance effects can
be observed at TRISTAN even if the colliding beam energy is
far away from the resonance peak,
due to the interference effects between the tensor
exchange amplitudes and the SM ones.
The experiments should hence be sensitive to the effects of
a tensor resonance of a much smaller leptonic width
or a much larger mass than has been postulated
as a possible origin of the 60\,GeV di-gamma events observed at LEP.

Finally, we found that the tensor boson exchange effects in the
processes $e^+ e^- \to \ell^+\ell^- (\ell \ne e)$ and $\gamma\gamma$
reduce to those of the contact higher-dimensional interactions
in the high tensor boson mass limit.
We identified the substitution rules between the tensor exchange
effects and the effects of dimension-6 contact $ee\ell\ell$
couplings\cite{eichten} for the process
$e^+e^- \to \ell^+ \ell^-$,
and those between the tensor exchange and the dimension-8
contact $ee\gamma\gamma$ interactions\cite{HH}
for the process $e^+e^- \to \gamma \gamma$.

\section*{Acknowledgements}

The authors wish to thank T.\,Kawai, M.\,Kobayashi, S.\,Komamiya,
T.\,Mori, Y.\,Okada, M.\,Tanabashi, and T.\,Tsukamoto
for interesting discussions.
We also wish to thank N.A.\,McDougall for a careful
reading the manuscript.


\newpage


\begin{figure}
\caption{
Feynman diagrams for the decay
$Z \to \phi_{{_S},{_P}} \ell_L \,\overline{\ell_R}$.
\label{fig.1}}
\end{figure}

\begin{figure}
\caption{
Dalitz distributions (a) $\Sigma_V$ and (b) $\Sigma_A$ of eqs.(2.7)
and (2.8), respectively, for $m=59$\,GeV.
\label{fig.2}}
\end{figure}

\begin{figure}
\caption{
Feynman diagrams for the decay
$Z \to \phi_{_T} \ell_R \overline{\ell_R}$.
\label{fig.3}}
\end{figure}

\begin{figure}
\caption{
Dalitz distribution $\Sigma_T$ of eq.(2.26) for the decay
$Z \to \phi_{_T} \ell_R \, \overline{\ell_R}$
\label{fig.4}}
\end{figure}

\begin{figure}
\caption{
The energy dependence of the differential cross section
$d\sigma(e^+e^- \to e^+e^-)/d\cos\theta$ at
$\cos\theta=0$, 0.4 and 0.8 for $f_{_R}^e = 0.0086$,
$f_{_L}^e = 0$ and $\Gamma_{\!_T} = 100 $\,MeV.
The dashed lines give the SM predictions for
$e^2 =4\pi\overline{\alpha}(m^2_{_Z})$ with
$\overline{\alpha}(m^2_{_Z}) =1/128$, $\sin^2\theta_W =0.23$,
$m_{_Z} =91.187$GeV and $\Gamma_{\!_Z}=2.5$GeV.
\label{fig.5}}
\end{figure}

\begin{figure}
\caption{
The energy dependence of the differential cross section
$d\sigma(e^+e^- \to \mu^+\mu^-)/d\cos\theta$ at
$\cos\theta=0$ for $f_{_R}^e = \pm f_{_R}^\ell= 0.0086$
and $f_{_L}^e = f_{_L}^\ell = 0$.
The dashed line gives the SM prediction for the SM parameters
of Fig.~5.
\label{fig.6}}
\end{figure}

\begin{figure}
\caption{
The energy dependence of the differential cross section
$d\sigma(e^+e^- \to \gamma\gamma)/d\cos\theta$
at $\cos\theta=0$
for $f_{_R}^e = 0.0086$, $f_{_L}^e = 0$ and $h= \pm 0.10$.
The dashed line gives the SM prediction for
$e^2 =4\pi\alpha$ with $\alpha=1/137$.
\label{fig.7}}
\end{figure}

%
%

\end{document}